\documentstyle[prl,floats,epsbox,twocolumn,aps]{revtex}

\begin{document}
\draft

\twocolumn[\hsize\textwidth\columnwidth\hsize\csname
@twocolumnfalse\endcsname

\title{ \bf Coupled-Cluster Approach to Electron Correlations in \\
the Two-Dimensional Hubbard Model}
\author{Yoshihiro Asai and Hideki Katagiri}
\address{Electrotechnical Laboratory (ETL), \\
Agency of Industrial Science and Technology (AIST), \\
Umezono 1-1-4, Tsukuba, Ibaraki 305, Japan}
\date{Received on \today}
\maketitle

\widetext

\begin{abstract}
We have studied electron correlations in the doped two-dimensional (2D) 
Hubbard model by using the coupled-cluster method (CCM) to investigate
whether or not the method can be applied to correct the independent
particle approximations actually used in {\it ab-initio} band
calculations. The double excitation version of the CCM, implemented using
the approximate coupled pair (ACP) method, account for most of the
correlation energies of the 2D Hubbard model in the weak ($U/t \simeq 1$)
and the intermediate $U/t$ regions ($U/t \simeq 4$). The error is always
less than $1 \%$ there. The ACP approximation gets less accurate for
large $U/t$ ($U/t \simeq 8$) and/or near half-filling. Further
incorporation of electron correlation effects is necessary in this region.
The accuracy does not depend on the system size and the gap between the
lowest unoccupied level and the highest occupied level due to the finite
size effect. Hence, the CCM may be favorably applied to {\it ab-initio}
band calculations on metals as well as semiconductors and insulators.
\end{abstract}

]

\narrowtext

Electron correlation in solids is the origin of various macroscopic
quantum phenomena such as magnetism and high-$T_c$ 
superconductivity.~\cite{Fulde,AsaiO}
Interests in these fields are gradually shifting from materials
which have simple chemical compositions and structures to more 
complicated ones. Besides the importance of simplified models such as the
Hubbard and Heisenberg models in statistical physics, the importance of
approximate many-body theories which can be applied to {\it ab-initio}
calculations~\cite{CrysHF,CrysAP} is rapidly increasing.       

The coupled cluster method (CCM) has been recognized as one of the most 
successful approximate methods for quantum many-body problems in nuclear
physics~\cite{NuclearPhys} and {\it ab-initio} quantum 
chemistry.~\cite{QuantChem} The method has also been successfully applied
to the electron gas problem,~\cite{Gas} but there have been few 
applications to quantum lattice problems. Within the limited number of
examples in the literature, we find that the method was successfully
applied to quantum spin  models,~\cite{SpinM} the one-dimensional Hubbard
and Pariser-Parr-Pople (PPP) model,~\cite{Hubb_Paldus} and the
two-dimensional (2D) Hubbard model at half-filling and with one 
hole.~\cite{HubbM} These successful results suggest the CCM may be one of
the most suitable methods to study electron correlations in solids as well
as molecules and nuclei.

To verify this expectation, we have made more rigorous tests of
the CCM than the previous ones which were made on the one-dimensional
Hubbard model on small clusters.~\cite{Hubb_Paldus} We studied the doped
2D Hubbard model at closed shell fillings on finite size square lattices
up to the $8 \times 8$ lattice by using the CCM. The total energies were
compared with those obtained by using the projector auxiliary field
quantum monte carlo (PAFQMC) method and the the adaptive sampling quantum
monte carlo (ASQMC) method.~\cite{ASQMC} (An augmented version of the
PAFQMC method which reduces the difficulty of the negative sign problem.)
We have studied how the accuracy of the CCM depend on the Fermi degeneracy
or magnitudes of the gaps of the system to be studied as well as
magnitudes of Coulomb interactions and the electron fillings. The former
was not studied in the previous work, but it is very crucial if one tries
to use the CCM for solid state electrons.

We have studied the 2D Hubbard model:
$
H =
-t \sum_{\langle i,j \rangle \sigma}
( c^\dagger_{i \sigma} c_{j \sigma} + H.C. )
+ U \sum_{i} n _{i \uparrow} n _{i \downarrow}.
$
We put $t=1$.

In the CCM, the exact totally symmetric nondegenerate ground state 
wavefunction $| \Psi \rangle$ is expressed on the basis of the exponential
Ansatz:
\begin{displaymath}
| \Psi \rangle = \exp (T) | \Phi \rangle ,
\end{displaymath}
$T = \sum_n T_n$ , where $T_n$ is a n-body excitation cluster operator.
If $\Phi$ is the Hartree-Fock (HF) wavefunction obtained in the
Hartree-Fock limit (easily obtained for a lattice Fermion model), we may
neglect the single excitation cluster $T_1$. The double excitation
cluster may then be sufficient: $T \approx T_2$ . This is an approximation
called the coupled-cluster-double (CCD) approximation.
Further incorporation of electron correlation is possible by taking into
account of the remaining clusters $T_3, T_4, \cdots$. It should be noted
however that owing to the exponential form: $\exp (T)$, higher excited
configurations than doubly excited configuration, such as quadruple
excitation etc., are taken into account in the CCD approximation.
If we use the normal product form of the Hamiltonian $H_N $defined in
time-independent perturbation theory,~\cite{t-indep} the Schr\"{o}dinger
equation may be written as follows:
\begin{displaymath}
( H_N  - \Delta E ) \exp (T) | \Phi \rangle = 0 ,
\end{displaymath}
where $\Delta E = E - E_{HF}$ is the difference of the exact ground state
energy and the Hartree-Fock energy. The Schr\"{o}dinger equation is solved
by multiplying $\exp (-T)$ from the left hand side of the former equation
and projecting the resultant equation on excited configurations 
made from $\Phi$:~\cite{QuantChem}
\begin{displaymath}
\langle \Phi_{ij}^{ab} | [H_N \exp (T) ]_C | \Phi \rangle = 0 ,
\end{displaymath}
where the subscript C indicates that only connected diagrams are taken 
into accounted. The total energy $E$ is given by:
\begin{displaymath}
E = E_{HF} + \langle \Phi | [H_N \exp (T) ]_C | \Phi \rangle .
\end{displaymath}
The former equations are the coupled-cluster equations and they and the
latter equation constitute a sufficient condition to the Schr\"{o}dinger
equation in the projected space. The advantage of the CCM is that $\exp
(T)$ is exactly manipulated  without artificial truncations owing to the
presence of the subscript C, unlike other variational theories. This
guarantees exactly the size consistency condition, which is difficult to
satisfy in  variational theories. However, there is no variational upper
bound given by the CCM energy. The double excitation cluster is expressed
as:
\begin{displaymath}
T_2 = 1/4 \sum_{i,j,a,b} t_{ij}^{ab} a^\dagger i b^\dagger j ,
\end{displaymath}
where $i$ and $j$ are annihilation operators for single electron states
occupied in $\Phi$ and $a^\dagger$ and $b^\dagger$ are creation operators
for single electron states unoccupied in $\Phi$.
The CCD coupled cluster equation is given by:
$
(\epsilon_i + \epsilon_j - \epsilon_a - \epsilon_b) t_{ij}^{ab} = 
\langle ij || ab \rangle +
1/2 \sum_{cd} \langle ab || cd \rangle t_{ij}^{cd} +
1/2 \sum_{kl} \langle ij || kl \rangle t_{kl}^{ab} - 
\sum_{kd} 
( \langle bk || jd \rangle t_{ki}^{ad} - \langle bk || id \rangle t_{kj}^{ad}
- \langle ak || jd \rangle t_{ik}^{db} + \langle ak || id \rangle t_{jk}^{db} )
+ \sum_{klcd} \langle kl || cd \rangle
[   t_{ki}^{ac} t_{lj}^{bd} + t_{ki}^{bd} t_{lj}^{ac} -
1/2 ( t_{ki}^{ab} t_{jl}^{cd} + t_{jl}^{ab} t_{ik}^{cd} ) -
1/2 ( t_{ij}^{ac} t_{kl}^{bd} + t_{ij}^{bd} t_{kl}^{ac} ) +
1/4 t_{ij}^{cd} t_{kl}^{ab} ] . \nonumber
$
The total correlation energy is given by:
\begin{displaymath}
\Delta E =
1/4 \sum_{ijab} \langle ij || ab \rangle
( t_{ij}^{ab} + t_i^a t_j^b - t_i^b t_j^a ) .
\end{displaymath}
$ \langle ij || ab \rangle $ is the anti-symmetrized two-electron
integral defined by:
$ 
\langle ij || ab \rangle = \langle ij | ab \rangle - \langle ij | ba \rangle .
$
$\epsilon_i$ is the i-th single particle energy.
Each term in the CCD coupled-cluster equations can be assigned to a
diagram. It is known based on diagram theoretical arguments that the
effect of the quadruple cluster $T_4$ may be approximately taken into
account by neglecting the 8, 9, 13, and 14-th terms in the right hand side
of the CCD coupled-cluster equations. Such an approximation is known as
the approximate coupled pair (ACP) 
approximation.~\cite{Hubb_Paldus,ACP_Paldus}

\begin{figure}[htbp]
\begin{center}
\epsfxsize=8.5cm
\epsffile{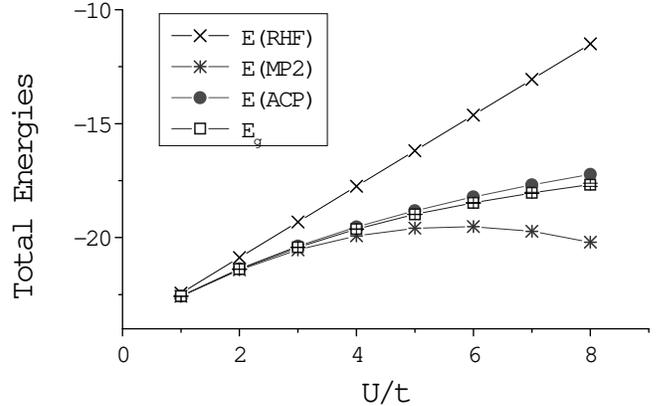}
\end{center}
\vspace*{1.0cm}
\caption{ Total energies of the $4 \times 4$ Hubbard model calculated
by using some approximations. $E_g$ denotes the exact ground state energy
calculated with the QMC method. $\rho = 0.625$.
The energies are plotted as a function of $U/t$.
Notations for the symbols are summarized in the legends.}
\label{fig: eng4x4}
\end{figure}

We have calculated the total energy of the 2D Hubbard model by using the
ACP approximation. We have compared the ACP energies with exact energies
obtained by using the PAFQMC and/or ASQMC methods. These are simply
denoted by the QMC method, hereafter. The calculations were made on the $4
\times 4$, $6 \times 6$, and $8 \times 8$ lattices. Electron fillings
$\rho = N_e / N_s$ are such that electrons form closed shells on these
finite size lattices, where $N_e$ is the number of electrons and $N_s$ is
the number of sites. We did not use the second order perturbative
selection rules frequently used in {\it ab-initio} quantum
chemistry calculations, which raise systematic errors in results when
$U/t$ is large. We have improved an existing fast algorithm by utilizing
symmetries of two-electron integrals in the Hubbard model.~\cite{FastALG} 
It takes about 10 minutes of CPU time for a calculation of the
$8 \times 8$ lattice on the Alpha workstation with the Alpha 21164/533 MHz
CPU chip when we use the ACP approximation. It would not be difficult to
do a CCM calculation even beyond a $128$ site lattice. We have not done
this because the QMC calculations take much more CPU time than the ACP
calculations.

\begin{figure}[htbp]
\begin{center}
\epsfxsize=8.5cm
\epsffile{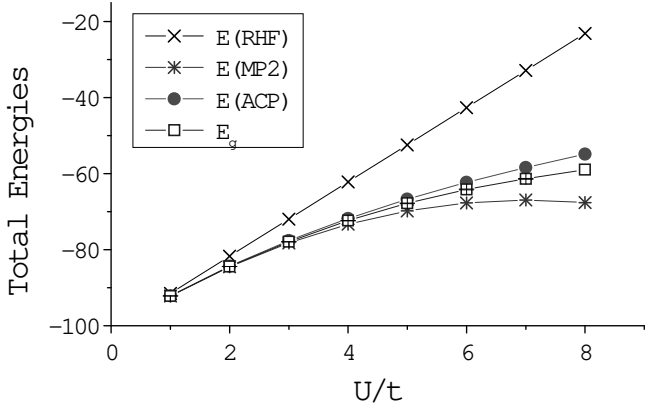}
\end{center}
\vspace*{1.0cm}
\caption{ Total energies of the $8 \times 8$ Hubbard model calculated
by using some approximations. $E_g$ denotes exact ground state energy
calculated with the QMC method. $\rho = 0.78125$.
The energies are plotted as a function $U/t$.
Notations for the symbols are summarized in the legends.}
\label{fig: eng8x8}
\end{figure}

The ground state energies of the 2D Hubbard model on the $4 \times 4$
lattice calculated with various methods are plotted in 
Fig. \ref{fig: eng4x4}.
The electron filling $\rho$ is $0.625$. The Restricted Hartree-Fock (RHF)
method gives very poor values of the ground state energy even for very
small values of $U/t$. The second order Moller-Plesset (MP2) approximation
works fairly well in the small $U/t$ region ($U/t \simeq 1)$, but it fails
in the intermediate ($U/t \simeq 4$) and the large $U/t$ regions ($U/t
\simeq 8$). The results obtained by using the ACP approximation are in
good agreement with the QMC energies up to the intermediate $U/t$ region
on this lattice. The similar results obtained on the $8 \times 8$ lattice
are plotted in Fig. \ref{fig: eng8x8}. $\rho$ is $0.78125$. Again, the RHF
method is a very poor approximation even for very  small value of $U/t$. 
The MP2 approximation works only in the small $U/t$ region. The ACP
approximation works well up to the intermediate $U/t$ region. The
deviation of the energy calculated by using the ACP approximation from the
energy calculated by using the QMC method on the $8 \times 8$ lattice is
somewhat larger than that obtained on the $4 \times 4$ lattice when
$U/t=8$.

\begin{figure}[htbp]
\begin{center}
\epsfxsize=8.5cm
\epsffile{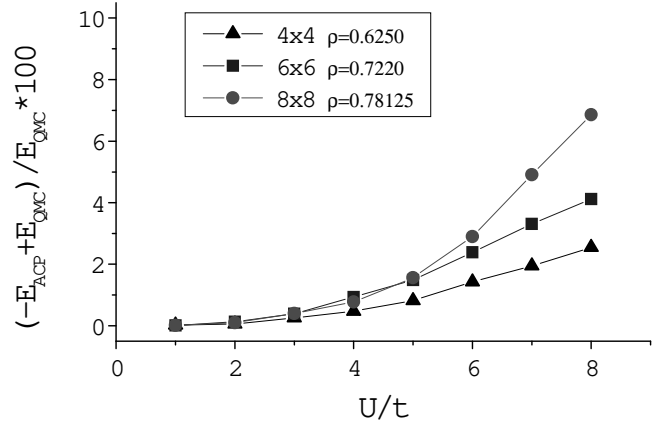}
\end{center}
\vspace*{1.0cm}
\caption{ $U/t$ dependencies of errors (\%) of the ACP approximation,
calculated on $4 \times 4$, $6 \times 6$, and $8 \times 8$ lattices.        
The error (\%) is defined by $(-E_{ACP}+E_{QMC})/E_{QMC} \times 100$.
$\rho$ for the $4 \times 4$, $6 \times 6$, and $8 \times 8$ lattices
are, $0.6250$, $0.7220$, and $0.78125$, respectively.
Notations for the symbols are summarized in the legends.}
\label{fig: error}
\end{figure}

The errors $\%$ of the ACP approximation defined by
$(- E_{ACP} + E_{QMC})/E_{QMC} \times 100$ calculated on $4 \times 4$, $6
\times 6$, and $8 \times 8$ lattices are plotted against $U/t$ in Fig.
\ref{fig: error}. $\rho$ of the $4 \times 4$, $6 \times 6$, and 
$8 \times 8$ lattices are $0.6250$, $0.7220$, and $0.78125$, respectively.
In all cases studied, the errors are less than $1 \%$, when $U/t \le 4$.
The errors grow as we increase $U/t$ and they are larger than $2 \%$ when
$U/t=8$. The error in the $8 \times 8$ lattice reaches almost $7 \%$ when
$U/t = 8$. Thus the ACP approximation breaks down in the large $U/t$
region. The errors increase most rapidly in the case of the $8 \times 8$
lattice. The enhancement of the error in the larger lattice may be mostly
due to the difference of $\rho$. $\rho$ is closest to $1$ in the $8
\times 8$ lattice in our cases. However, it is interesting to ask
whether or not the enhancement is brought about by the decrease of the gap
between the lowest unoccupied level (LUL) and the highest occupied level
(HOL) in the finite size cluster.

\begin{figure}[htbp]
\begin{center}
\epsfxsize=8.5cm
\epsffile{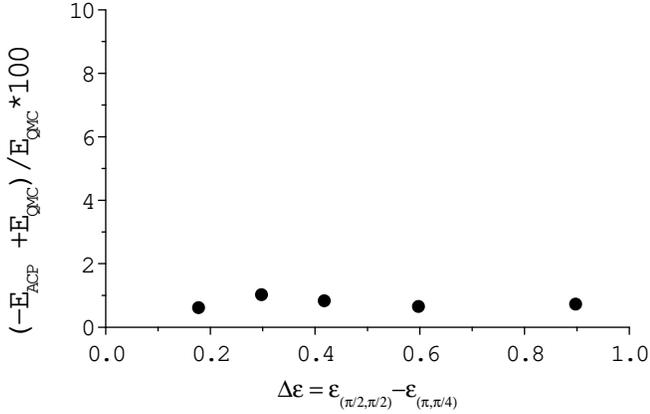}
\end{center}
\vspace*{1.0cm}
\caption{ The error (\%) $(-E_{ACP}+E_{QMC})/E_{QMC} \times 100$
as a function of $\Delta \epsilon$ in the generalized 2D Hubbard model 
on the $8 \times 8$ lattice.}
\label{fig: genHubb}
\end{figure}

To clarify this point, we study the following generalized
Hubbard model:
$
H =
-t \sum_{\langle i,j \rangle \sigma}
( c^\dagger_{i \sigma} c_{j \sigma} + H.C. )
+ U \sum_{i} n _{i \uparrow} n _{i \downarrow}
-t' \sum_{(i,j) \sigma}
(c^\dagger_{i \sigma} c_{j \sigma} + H.C. )
-t'' \sum_{[i,j] \sigma}
(c^\dagger_{i \sigma} c_{j \sigma} + H.C. ), 
$
where $(i,j)$ and $[i,j]$ denote the second and third nearest pairs of
sites on the square lattice, respectively. We put $t=1, t'=-0.2$, and
$U=4$. $t''$ is a variable. We introduced anisotropy $\pm 0.0001$ on the
$x$ and $y$ components of $t$ and $t''$, as well as on the $(1,1)$ and
$(1,-1)$ components of $t'$. We studied the $8 \times 8$ lattice system
with $78$ electrons. If $0.0035 \le t''$ the wave number of the HOL is
$(\pi, \pi/4)$ and the wave number of the LUL is $(\pi/2, \pi/2)$. The
energy separation  
$
\Delta \epsilon \equiv  \epsilon_{(\pi/2, \pi/2)}-\epsilon_{(\pi, \pi/4)}
$ 
is a monotonically increasing function of $t''$. We plot the errors $\%$
of the ACP approximation $(- E_{ACP} + E_{QMC})/E_{QMC} \times 100$ as a
function of $\Delta \epsilon$ in Fig. \ref{fig: genHubb}. The error is 
almost independent of $\Delta \epsilon$ and is always less than $1 \%$, 
as long as the CCM equations converge. The CCD equations with the ACP
approximation do not converge when $\Delta \epsilon \le 0.18$. 
This $\Delta \epsilon$ dependence of the errors indicates the enhancements
of the errors in the larger lattices observed in Fig. \ref{fig: error}
comes from the difference of the electron fillings $\rho$. The CCM is less
accurate close to the half-filling.

While it seems the ACP approximation works nicely in the Hubbard model
up to the intermediate $U/t$ region, it fails in the large $U/t$ region  
and/or close to half-filling in the two dimension. The $U/t$
dependence of the error was also observed in one-dimensional Hubbard model
on small clusters. Direct incorporation of $T_n, n \ge 3$ may be necessary
in the large $U/t$ region and close to the half-filling. The point in this
article is that the accuracy of the CCM does not depend much on the
LUL-HOL gap $\Delta \epsilon$. When we apply the crystal orbital method
to solid electrons, we use a finite number of $k$ points for $k$ space
integrations and constructing excited configurations.~\cite{CrysHF,MP2}
Hence, the Fermi degeneracy in metals is replaced by a pseudo-degeneracy
among some of the $k$ points. The situation is similar to atomic cluster
calculations. In numerical calculations, there are implicitly gaps due to
the finite number of the $k$ points, even for metallic electrons, as well
as for electrons in semiconductors and insulators. The fact that the
accuracy of the CCM does not depend on such gaps is very favorable, unless
numerical results are very dependent on details of the calculation
parameters such as increments of $k$ rather than on physical parameters.
Hence, the CCM may be successfully applied to {\it ab-initio} band
calculations on metals as well as semiconductors and insulators.

In conclusion, we have studied the electron correlation problem in the
doped 2D Hubbard model with the ACP approximation to investigate whether
or not the approximation can be applied to correct the independent
particle approximations really used in {\it ab-initio} band calculations. 
We found most of correlation energies of the 2D Hubbard model in the weak
and the intermediate $U/t$ region can be accounted for by the ACP
approximations.  In the large $U/t$ region and/or close to the
half-filling, the error of the ACP approximation is not negligible.
Further incorporation of the electron correlation effect may be necessary
there. The accuracy of the CCM does not depend on the LUL-HOL gap. Hence,
the CCM may be successfully applied to {\it ab-initio band} calculations
on metals as well as semiconductors and insulators.

One of the authors (Y.A.) appreciates Prof. B.A. Friedman for reading
this manuscript and giving some comments.
This work is supported financially by the project E-TK980204 in ETL.
Part of calculations were made by using facilities of
Research Information Processing System (RIPS) center of AIST and
the Supercomputer Center, Institute for Solid State Physics (ISSP),
University of Tokyo, to which the authors would like to express thanks.


%
%
%
%

\end{document}